\documentclass[10pt,twoside]{article}
\usepackage{amssymb}
\usepackage{graphicx}
\usepackage{amsmath}
\usepackage{psfrag}
\usepackage{feynarts}
\usepackage[latin1]{inputenc}

\begin{document}

\title{Improving bounds on flavor changing vertices in the two Higgs doublet
model from $B^{0}-\bar{B}^{0}$ mixing.}
\author{Rodolfo A. Diaz$^1$\thanks{%
radiazs@unal.edu.co}, R. Martinez$^1$\thanks{%
remartinezm@unal.edu.co}, Carlos E. Sandoval$^2$\thanks{%
cesandovalu@unal.edu.co} \\
1.Universidad Nacional de Colombia, \\
Departamento de Física. Bogotá, Colombia\\
2. Universität Hamburg, II. Institut für Theoretische Physik,\\
Luruper Chaussee 149, 22761, Hamburg, Germany}
\date{}
\maketitle

\vspace{-7mm}

\begin{abstract}
We find some constraints on the flavor changing vertices of the two Higgs
doublet model, from the $\Delta M_{B_{d}}$ measurement. Although bounds from
this observable have already been considered, this paper takes into account
the role of a new operator not included previously, as well as the vertices $%
\xi _{bb},\ \xi _{tc}$ and $\xi _{sb}$. Using the Cheng-Sher
parametrization, we found that for a relatively light charged Higgs boson
(200-300 GeV), we get that $\left\vert \lambda _{tt}\right\vert \lesssim 1$,
while the parameter $\lambda _{bb}$ could have values up to about 50. In
addition, we use bounds for $\lambda _{tt}$ and $\lambda _{bb}$ obtained
from $B^{0}\rightarrow X_{s}\gamma $ at next to leading order, and studied
the case where the only vanishing vertex factors are the ones involving
quarks from the first family. We obtained that $\Delta M_{B_{d}}$ is not
sensitive to the change of the parameter $\lambda _{sb}$, while $\left\vert
\lambda _{tc}\right\vert \lesssim 1$
\end{abstract}

The simplest extension of the SM compatible with gauge invariance is the so
called Two Higgs Doublet Model (2HDM), in which the second Higgs doublet is
identical to the SM one \cite{rdiaz}. In this model, the particle spectrum
is enlarged by the appearence of five Higgs bosons, two of them neutral
CP-even, a neutral CP-odd and two charged ones. A new feature of the 2HDM
consists of the appearence of processes with flavor changing neutral
currents (FCNC). One of the main motivations to study scenarios with FCNC is
the increasing evidence on neutrino oscillations that leads to lepton flavor
violation (LFV) \cite{R1}.\newline
In this paper we are concerned with FCNC in the quark sector in the
framework of the 2HDM type III, in which such processes are allowed at tree
level. Recently, constraints on the lepton and quark sectors have been found
from leptonic decays, $B$ meson decays and the $B^{0}-\bar{B}^{0}$ mixing 
\cite{sando1, xiao}. In Ref. \cite{sando1} the box diagrams are assumed
negligible while Ref. \cite{xiao} assumes the box diagrams to be dominant.
Notwithstanding, the latter reference does not include some operators and
vertices that could contribute to the box diagrams significantly. We intend
to study the effect of an operator and some vertices not considered in \cite%
{xiao}.

\section*{$\Delta M$ calculation}

\begin{figure}[tbp]
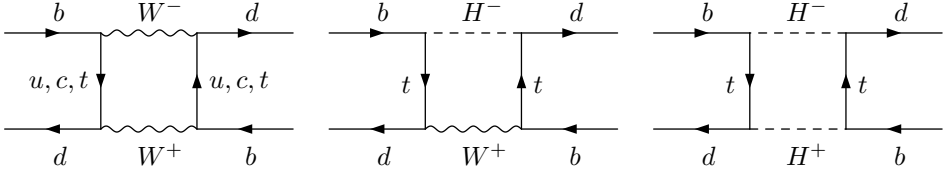

\begin{center}
\begin{feynartspicture}(300,130)(3,1)
\FADiagram{} 
\FAProp(-9.,10.)(-1.,10.)(0.,){/Straight}{1}
\FAProp(-1.,10.)(-1.,2.)(0.,){/Straight}{1}
\FAProp(-1.,2.)(-9.,2.)(0.,){/Straight}{1}
\FAProp(-1.,2.)(7.,2.)(0.,){/Sine}{0}
\FAProp(7.,2.)(7.,10.)(0.,){/Straight}{1}
\FAProp(7.,10.)(-1.,10.)(0.,){/Sine}{0}
\FAProp(7.,10.)(15.,10.)(0.,){/Straight}{1}
\FAProp(7.,2.)(15.,2.)(0.,){/Straight}{-1}
\FALabel(-5.,11.)[bl]{$b$}
\FALabel(-5.,-1.)[bl]{$d$}
\FALabel(-7.,5.)[bl]{$u,c,t$}
\FALabel(2.,11.)[bl]{$W^-$}
\FALabel(2.,-1.)[bl]{$W^+$}
\FALabel(11.,11.)[bl]{$d$}
\FALabel(11.,-1.)[bl]{$b$}
\FALabel(8.,5.)[bl]{$u,c,t$}
\FAProp(18.,10.)(26.,10.)(0.,){/Straight}{1}
\FAProp(26.,10.)(26.,2.)(0.,){/Straight}{1}
\FAProp(26.,2.)(18.,2.)(0.,){/Straight}{1}
\FAProp(26.,2.)(34.,2.)(0.,){/Sine}{0}
\FAProp(34.,2.)(34.,10.)(0.,){/Straight}{1}
\FAProp(34.,10.)(26.,10.)(0.,){/ScalarDash}{0}
\FAProp(34.,10.)(42.,10.)(0.,){/Straight}{1}
\FAProp(34.,2.)(42.,2.)(0.,){/Straight}{-1}
\FALabel(22.,11.)[bl]{$b$}
\FALabel(22.,-1.)[bl]{$d$}
\FALabel(24.,5.)[bl]{$t$}
\FALabel(29.,11.)[bl]{$H^-$}
\FALabel(29.,-1.)[bl]{$W^+$}
\FALabel(38.,11.)[bl]{$d$}
\FALabel(38.,-1.)[bl]{$b$}
\FALabel(35.,5.)[bl]{$t$}
\FAProp(45.,10.)(53.,10.)(0.,){/Straight}{1}
\FAProp(53.,10.)(53.,2.)(0.,){/Straight}{1}
\FAProp(53.,2.)(45.,2.)(0.,){/Straight}{1}
\FAProp(53.,2.)(61.,2.)(0.,){/ScalarDash}{0}
\FAProp(61.,2.)(61.,10.)(0.,){/Straight}{1}
\FAProp(61.,10.)(53.,10.)(0.,){/ScalarDash}{0}
\FAProp(61.,10.)(69.,10.)(0.,){/Straight}{1}
\FAProp(61.,2.)(69.,2.)(0.,){/Straight}{-1}
\FALabel(49.,11.)[bl]{$b$}
\FALabel(49.,-1.)[bl]{$d$}
\FALabel(51.,5.)[bl]{$t$}
\FALabel(56.,11.)[bl]{$H^-$}
\FALabel(56.,-1.)[bl]{$H^+$}
\FALabel(65.,11.)[bl]{$d$}
\FALabel(65.,-1.)[bl]{$b$}
\FALabel(62.,5.)[bl]{$t$}
\end{feynartspicture}
\end{center}
\caption{{\protect\small {Box diagrams for $B^{0}-\bar{B}^{0}$ in the 2HDM.}}
}
\label{box}
\end{figure}
The relevant Feynman diagrams for this process are shown in figure \ref{box}%
. The calculation for the SM was first performed in \cite{inami}, where the
diagrams involving gauge bosons are changed by diagrams with Goldstone
bosons $\phi ^{\pm }$ considering them as with the same mass of the $W$. The
expression for $\Delta M$ in the framework of the SM reads \cite{buras3}: 
\begin{equation*}
\Delta M_{B_{d}}=\frac{G_{F}^{2}}{6\pi ^{2}}%
m_{B}|V_{td}V_{tb}|^{2}B_{B}f_{B}^{2}m_{W}^{2}\eta _{B}S_{0}(x_{t}),
\end{equation*}%
where 
\begin{eqnarray}
S_{0}(x_{wf}) &=&\frac{4x_{wf}-11x_{wf}^{2}+x_{wf}^{3}}{4(1-x_{wf})^{2}}-%
\frac{3x_{wf}^{3}}{2(1-x_{wf})^{3}}\log (x_{wf}). \\
x_{ij} &\equiv &\left( \frac{m_{j}}{m_{i}}\right) ^{2}
\end{eqnarray}%
The $B_{B}$ and $\eta _{B}$ functions are the non-perturbative and
perturbative QCD corrections respectively. Finally, $f_{B}$ refers to the
decay constant of the $B$ meson. On the other hand, regarding the extended
Higgs sector, the calculation on 2HDM of type I and II and a study of $%
\Delta M$ including QCD corrections was made in \cite{urban}, and for the
model type III $\Delta M\ $was studied in \cite{xiao}.

\begin{figure}[tbh]
\begin{center}
\psfrag{fb}{$f_B\sqrt{B_B}$} \psfrag{lambdatc}{$\lambda_{tc}$} %
\includegraphics[height=4.445cm]{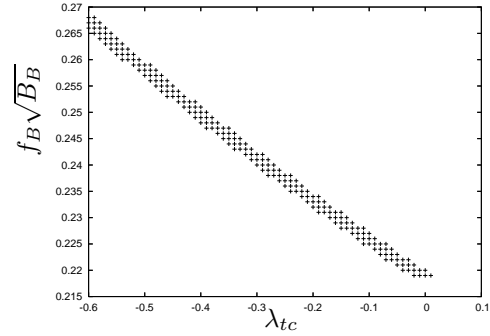}
\end{center}
\caption{Contourplot on the $f_{B}\protect\sqrt{B}_{B}-\protect\lambda _{tc}$
plane with $\protect\lambda _{sb}=0$, $m_{H}=250$ GeV, taking $\protect%
\lambda _{tt}=0.5,\protect\lambda _{bb}=22$.}
\label{contour}
\end{figure}

In order to make this calculation in the framework of the 2HDM type III, we
shall make the following approximations 1) For two identical quarks in the
loop, we shall only take into account the contribution due to the top quark.
2) We shall consider that FC vertices $\xi _{ij}\ $involving the first
generation are negligible. Combining both approximations, we find that the
coefficients $R_{tq}^{U,D}\ $in the Yukawa Lagrangian should be taken as%
\begin{eqnarray}
R_{td}^{D} &=&0\ \ ;\ \ R_{tb}^{D}=V_{ts}\xi _{sb}+V_{tb}\xi _{bb}  \notag \\
R_{td}^{U} &=&\xi _{tc}V_{cd}+\xi _{tt}V_{td}\ \ ;\ \ R_{tb}^{U}=\xi
_{tc}V_{cb}+\xi _{tt}V_{tb}  \label{Rtq}
\end{eqnarray}%
we shall also use the Cheng-Sher parametrization for the FC vertices%
\begin{equation}
\xi _{qq^{\prime }}=\left( \sqrt{2}G_{F}m_{q}m_{q^{\prime }}\right)
^{1/2}\lambda _{qq^{\prime }}  \label{ltq}
\end{equation}%
and the contributions for$\ \Delta M_{B_{d}}\ $read 
\begin{equation*}
\Delta M_{B_{d}}=\frac{G_{F}^{2}}{6\pi ^{2}}(V_{td}^{\dag
}V_{tb})^{2}B_{B}f_{B}^{2}\eta _{B}m_{B}^{2}m_{W}^{2}S_{2HDM},
\end{equation*}%
where 
\begin{eqnarray}
S_{2HDM} &=&S_{0}(x_{wt})+S_{HH}(x_{H}(m_{t}))\left( \sqrt{\frac{m_{c}}{m_{t}%
}}\frac{V_{cd}}{V_{td}}\frac{\lambda _{tc}}{\lambda _{tt}}+1\right)
^{2}\left( \sqrt{\frac{m_{c}}{m_{t}}}\frac{V_{cb}}{V_{tb}}\frac{\lambda _{tc}%
}{\lambda _{tt}}+1\right) ^{2}  \notag \\
&-&5\frac{m_{B}^{2}}{(m_{b}+m_{d})^{2}}S_{HH}^{\prime }(x_{H}(m_{t}))\left( 
\sqrt{\frac{m_{c}}{m_{t}}}\frac{V_{cd}}{V_{td}}\frac{\lambda _{tc}}{\lambda
_{tt}}+1\right) ^{2}\left( \sqrt{\frac{m_{s}}{m_{b}}}\frac{V_{ts}}{V_{tb}}%
\frac{\lambda _{sb}}{\lambda _{bb}}+1\right) ^{2}  \notag \\
&+&S_{WH}(x_{H}(m_{t}),x_{W}(m_{t}))\left( \sqrt{\frac{m_{c}}{m_{t}}}\frac{%
V_{cd}}{V_{td}}\frac{\lambda _{tc}}{\lambda _{tt}}+1\right) \left( \sqrt{%
\frac{m_{c}}{m_{t}}}\frac{V_{cb}}{V_{tb}}\frac{\lambda _{tc}}{\lambda _{tt}}%
+1\right) ,  \label{S2HDM}
\end{eqnarray}%
and 
\begin{eqnarray}
S_{HH}(x_{Ht}) &=&\lambda _{tt}^{4}\frac{x_{Ht}x_{Wt}}{4}\left( \frac{%
1+x_{Ht}}{(1-x_{Ht})^{2}}+\frac{2x_{Ht}\log (x_{Ht})}{(1-x_{Ht})^{3}}\right)
, \\
S_{WH}(x_{Ht},x_{Wt}) &=&\lambda _{tt}^{2}\frac{x_{Ht}x_{Wt}}{4}\left[ \frac{%
(2x_{Wt}-8x_{Ht})\log (x_{Ht})}{(1-x_{Ht})^{2}(x_{Ht}-x_{Wt})}\right.  \notag
\\
&&\left. +\frac{6x_{Wt}\log (x_{Wt})}{(1-x_{Ht})^{2}(x_{Ht}-x_{Wt})}-\frac{%
8-2x_{Wt}}{(1-x_{Ht})(1-x_{Wt})}\right] ,
\end{eqnarray}%
\begin{equation*}
S_{HH}^{\prime }(x_{Ht})=\lambda _{tt}^{2}\lambda _{bb}^{2}\frac{%
x_{Ht}x_{Hb}x_{Wt}}{4}\left( \frac{2(1-x_{Ht})+\log (x_{Ht})(1+x_{Ht})}{%
(1-x_{Ht})^{3}}\right)
\end{equation*}%
The function $S_{HH}^{\prime }$ comes from the vertex $\xi _{bb}$ and it was
not considered in Ref. \cite{xiao}. We have also taken into account the
perturbative QCD correction $\eta _{B}\ $taken from \cite{xiao}. The factor $%
f_{B}\sqrt{B}_{B}$, introduces a lot of uncertainity in most of the
calculations. In \cite{xiao}, one can find an estimate of this uncertainity,
obtained by plotting $V_{td}-f_{B}\sqrt{B}_{B}$, based on the experimental
value of $\Delta M$, obtaining allowed values between $0.19$ GeV and $0.27$ GeV. A
more stringent range between $0.219$ GeV and $0.273$ GeV is obtained from \cite%
{lepton-photon}, which will be the values we use in our analyses.\newline
Taking $\lambda _{bb}=0$, the results are the same as in \cite{xiao}, i.e,
it is concluded that $\lambda _{tt}$ should be less than one. On the other
hand, values greater than 0.7 would not be favored if one expects the
charged Higgs boson to be relatively light i.e in the region of $200-300$
GeV (we shall assume the charged Higgs boson to be relatively light
throughout the document). Adding the contribution of the $\lambda _{bb}$
factor, we find that for values between 30 and 50 of this vertex (which are
allowed by the $B\rightarrow X_{s}\gamma $ process \cite{xiao}), the maximum
values of $\lambda _{tt}$ could be lower than in the latter case. Finally,
it worths saying that these bounds are compatible with the ones imposed to $%
\lambda _{bb},\ \lambda _{tt}$ from perturbativity grounds \cite{Rozo}.%
\newline
Up to now we have considered that only the vertices $\lambda _{tt}$ and $%
\lambda _{bb}$ contributes to the process. Now, we shall study the
possibility of including the contributions of $\lambda _{tc}$ and $\lambda
_{bs}$ (not considered in Ref. \cite{xiao}). In that case, the coefficients $%
R_{tq}^{U,D}$ described in Eqs. (\ref{Rtq}, \ref{ltq}) should be taken in
complete form (but maintaining the approximations that led to Eq. (\ref{Rtq}%
)). We will use some of the restrictions found in \cite{xiao} for $\lambda
_{tt}$ and $\lambda _{bb}$ from the $B\rightarrow X_{s}\gamma $ process, to
reduce the number of free parameters and try to get new bounds on the new
parameters introduced. Taking $\lambda _{tt}=0.5$ and $\lambda _{bb}=22$, we
obtained that the behavior of $\Delta M$ as a function of $\lambda _{tc}$ is
basically independent of the value taken for $\lambda _{sb}$, at least by
assuming $\left\vert \lambda _{sb}\right\vert \leq 100$. The same occurs
when we took $\lambda _{tt}=0.5,\lambda _{bb}=1$. Since $\lambda _{sb}$
could take large values without affecting the behavior of $\Delta M$, it
would be useless to make a graph of $\Delta M$ as a function of this factor.%
\newline
On the other hand, by taking into account the big uncertainty in the $f_{B}%
\sqrt{B_{B}}$ factor, it could be interesting to see what region is
permitted by the experimental data for different values of $\lambda _{tc}$.
The results are shown in figure \ref{contour} for $\lambda _{tt}=0.5$ and $%
\lambda _{bb}=1$. The trend found in this part is somehow clear in what has
to do with $\lambda _{tc}$ and $\lambda _{sb}$. The vertex $\lambda _{tc}$
is the most constrained; together with $\lambda _{tt}$ are both less than
one, while $\lambda _{bb}$ and $\lambda _{sb}$ could have some higher
values. $\lambda _{bb}$ could be even 50, according to our results and to
the results in \cite{xiao}, while the values of $\lambda _{sb}$ do not
affect the function $\Delta M$ even for very large values.

Finally, there is a naive way to analize why $\Delta M$ is not sensitive to
the $\lambda _{sb}$ factor while it is for the $\lambda _{tc}$ vertex. By
taking the coefficients that accompany the operators $S_{HH}$ and $S_{WH}$
we can check that for values of $\left\vert \lambda _{tc}/\lambda
_{tt}\right\vert $ between $-1$ and $1$ we get regions in which the
contribution of $\lambda _{tc}$ is of the same order of the contribution of $%
\lambda _{tt}$ (in some cases constructive and in some cases destructive).
These contributions could also be significant for the new operator $%
S_{HH}^{\prime }$. By contrast, the quotient $\left\vert \lambda
_{sb}/\lambda _{bb}\right\vert $ should be at least of the order of 150 to
get a significant contribution from $\lambda _{sb}$ to the operator $%
S_{HH}^{\prime }$.

In conclusion, the combined data from $\Delta M_{B_{d}}\ $and\ $B\rightarrow
X_{s}\gamma $ could provide some information over the FC vertices $\lambda
_{bb},\ \lambda _{tt},\ \lambda _{tc},\lambda _{bs}$. A phenomenological
analysis shows that $\lambda _{bb}$ could still have large values up to
about $50$, the $\lambda _{sb}$ vertex keeps basically unconstrained while
the vertices $\lambda _{tt}\ $and$\ \lambda _{tc}$ are more restricted and
appears to be less than one in magnitude. 

\section{Acknowledgements}

We thank Colciencias, DINAIN, and HELEN for the financial support.

\end{document}